\documentclass[aps,prl,twocolumn,superscriptaddress]{revtex4}

\usepackage{graphicx}
\usepackage{epstopdf}
\usepackage{color}

\begin{document}


\title{Trapping molecular ions formed via photo-associative ionization of ultracold atoms}


\author{Scott T. Sullivan}
\affiliation{Department of Physics and Astronomy, University of California, Los Angeles, California 90095, USA}
\author{Wade G. Rellergert}
\affiliation{Department of Physics and Astronomy, University of California, Los Angeles, California 90095, USA}
\author{Svetlana Kotochigova}
\affiliation{Department of Physics, Temple University, Philadelphia, Pennsylvania 19122-6082, USA}
\author{Kuang Chen}
\affiliation{Department of Physics and Astronomy, University of California, Los Angeles, California 90095, USA}
\author{Steven J. Schowalter}
\affiliation{Department of Physics and Astronomy, University of California, Los Angeles, California 90095, USA}
\author{Eric R. Hudson}
\email{Eric.Hudson@ucla.edu}
\affiliation{Department of Physics and Astronomy, University of California, Los Angeles, California 90095, USA}

\date{\today}

\begin{abstract}
The formation of $^{40}$Ca$_2^+$ molecular ions is observed in a
hybrid $^{40}$Ca magneto-optical and ion trap system. The molecular
ion formation process is determined to be two-photon photo-associative
ionization of ultracold $^{40}$Ca atoms. A lower bound for the two-body,
two-photon rate constant is found to be $\bar{\beta} \geq 2 \pm 1
\times 10^{-15}$~cm$^{3}$~Hz. $\textit{Ab initio}$ molecular potential
curves are calculated for the neutral Ca$_2$ and ionic Ca$_2^+$ molecules 
and used in a model that identifies the photo-associative ionization pathway.
As this technique does not require a separate photo-association laser,
it could find use as a simple, robust method for producing ultracold,
state-selected molecular ions.

\end{abstract}


\maketitle

The low-energy internal structure of a diatomic molecule, $\textit{e.g.}$ the electric dipole moment and vibrational, rotational, and $\Omega$-doublet levels, presents a host of opportunities for advances in quantum simulation, precision measurement,  cold chemistry, and quantum information \cite{ColdMoleculesBook}. As such, the last decade has witnessed an enormous effort towards producing ultracold molecules in well-defined rovibrational states. It appears that this goal is in now within reach, as several groups have reported production of ultracold molecular samples in the rovibrational ground state \cite{NiKRB,Denschlag2008,Danzl2010}; in the case of the KRb work the sample is near quantum degeneracy \cite{NiKRB}.

To date, most work has focused on neutral molecules, however, a sample of ultracold molecular ions presents interesting possibilities. While the Coulomb interaction overwhelms the electric dipole interaction between polar molecular ions in most cases, many of the goals of cold molecule physics can be accomplished with molecular ions, with the added benefit of a simple, reliable trapping. Indeed, a cold sample of molecular ions allows the study of cold chemistry \cite{Willitsch2008}, which not only has important implications for understanding the formation of interstellar clouds \cite{smith_ion_1992}, but for investigation, and possible control, of reactive collisions in the quantum regime \cite{Hudson2006}; identification of carriers of the diffuse interstellar bands \cite{Reddy2010}; precision measurement of molecular transitions, which can be used to sensitively measure parity violating effects \cite{gottselig_mode-selective_2004}, as well as to constrain the possible variation of the fundamental constants \cite{flambaum_enhanced_2007}; and the implementation of a scalable quantum computation architecture \cite{andre_coherent_2006}.
\begin{figure}
\resizebox{1\columnwidth}{!}{
    \includegraphics{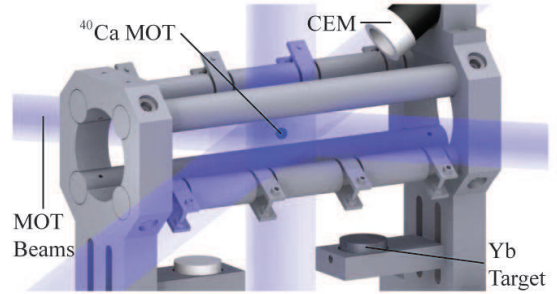}
}  \caption{MOTION trap:  A $^{40}$Ca magneto-optical trap is constructed inside a linear quadrupole ion trap.  Below the trap are two ablation targets used to load calibration ions.  The channel electron multiplier (CEM) is mounted above the trap. \label{motrender}}
\end{figure}
For these reasons, several groups have recently initiated work to produce samples of ultracold molecular ions \cite{Schneider2010,Staanum2010,Tong2010,Hudson2009}.  In Refs. \cite{Schneider2010,Staanum2010,Tong2010}, the molecular ion external degrees of freedom are sympathetically cooled with laser cooled atomic ions to ultracold temperatures. However, the long range ion-ion interaction does not couple to the molecular ion internal degrees of freedom and therefore the rovibrational temperature is unmodified by the sympathetic cooling process.  To overcome this limitation, optical pumping schemes \cite{Schneider2010,Staanum2010} and state selective photo-ionization of neutral molecules  \cite{Tong2010} have been used to produce molecular ions in the lowest few rotational states.  More generally, it should also be possible to use  ultracold $\textit{neutral}$ atoms,  in a magneto-optical trap (MOT), to simultaneously sympathetically cool both the internal and external degrees of freedom of molecular ions \cite{Hudson2009}.

In addition to these methods, it is possible to create state-selected ultracold molecular ions directly from their atomic constituents through photo-associative ionization (PAI), a two-body process where colliding neutral atoms are photo-excited into either a bound molecular ion or auto-ionizing molecular potential.  In this manuscript, we report the formation of $^{40}$Ca$_2^+$ molecular ions via PAI in a hybrid system composed of a $^{40}$Ca MOT constructed inside of a radio-frequency linear ion trap (MOTION trap). We quantify the production of the dimer ion using standard ion trap mass spectrometry techniques \cite{IonTrapping,Douglas2005} available for use in the MOTION trap system.  In the remainder of this work  we briefly describe the apparatus used and present measurements characterizing the molecular ion production rate. We conclude with results from $\textit{ab initio}$ calculations of the $^{40}$Ca$_2^+$ molecular ion structure and a description of the most-likely mechanism of molecular ion production.

\begin{figure}
\resizebox{1\columnwidth}{!}{
    \includegraphics{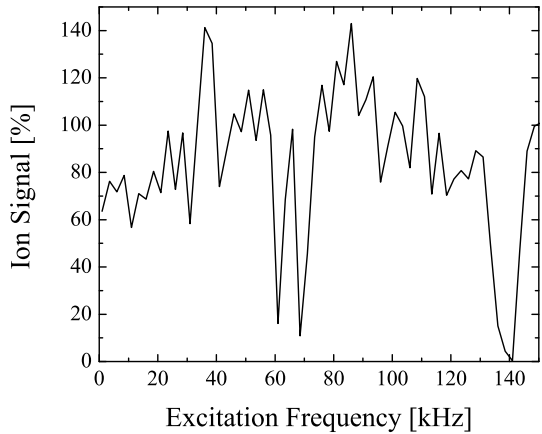}
}  \caption{Results of a excitation spectrum measurement performed at trapping parameters where the stability parameter (q) for m = 80~amu is 0.5, thus unstable for m = 40~amu.  The observed loss near 70~kHz is within 10\% of the predicted secular frequency for m = 80~amu.  The split peak feature is likely due to trap imperfections (asymmetric endcap placement, machining, imperfect quadrupole ratio,etc).    \label{secularscan}}
\end{figure}

	The MOTION trap system used in this work is that proposed in Ref. \cite{Hudson2009} and is shown in Fig. \ref{motrender}.  As aforementioned,  it is a hybrid system with a $^{40}$Ca MOT constructed inside an linear quadrupole ion trap (LQT), designed to explore collisions between ultracold atoms and trapped molecular ions. Briefly, cold $^{40}$Ca atoms are collected in a MOT by decelerating a beam of atoms produced by a Ca getter.  The $\lambda$ = 422.6~nm laser system , used to produce the MOT cooling and deceleration beams, is referenced to the 4$^{1}$P$_{1}$ $\leftarrow$ 4$^{1}$S$_{0}$ transition ($\Gamma = 2\pi \times 34.6$~MHz) using a saturated absorption signal from a Ca vapor cell.  The trapping (deceleration) laser beams typically have detuning and total  intensity of  $-\Gamma$ ($-3\Gamma$) and 23 (31)~mW/cm$^{2}$, respectively. The 671.7~nm $ 5^1$P$_1 \leftarrow 3^1$D$_2$ Ca repump transition is driven with a laser system referenced to the R69(2-5) molecular Iodine transition. The typical MOT atom number, density and temperature are measured by absorption and fluorescence imaging and found to be $N_{\rm{Ca}}$ = $1.0\pm 0.1\times10^7$~atoms, $\rho_{\rm{Ca}}$ = $3.9 \pm 0.5 \times 10^{10}$~cm$^{-3}$, and $T_{\rm{Ca}}$ = $4 \pm 1$~mK, respectively.  The LQT, which has a field radius of $r_o$ = 11.2~mm, is designed with three axial trapping regions to facilitate the shuttling of ions into and out of the center region for interaction with the MOT.  The number of trapped ions is detected by grounding the trap electrodes, allowing the ions to escape and be detected by a channel electron multiplier (CEM) ion detector located near the trap center.   The absolute ion number measured by the CEM is calibrated by comparing CEM detection to laser-induced fluorescence rates with $^{174}$Yb$^{+}$ ions loaded via laser ablation (see Fig. \ref{motrender}).  Only the central trapping region is operated in this experiment since the MOT acts as the source of all ions observed in this work.

\begin{figure}
\resizebox{1\columnwidth}{!}{
    \includegraphics{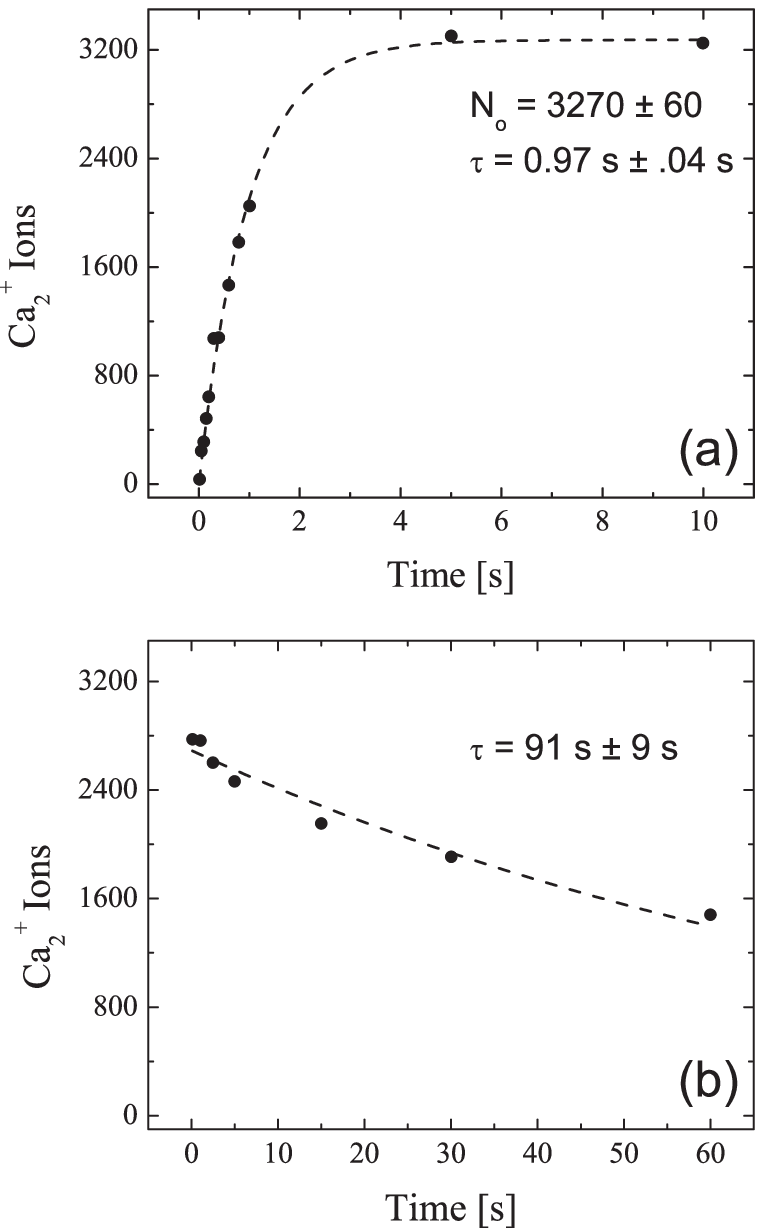}
}  \caption{Measurement of the ion number as a function of loading time is shown in panel (a).  In panel (b), trap lifetime is measured by loading the trap to saturation and stopping production for a fixed amount of time before detection.  Characteristic times are found by exponential fits (dashed lines).\label{Ca2+LoadingUnloading}}
\end{figure}

	When the MOTION system is operated without deliberately loading ions into the LQT, an accumulation of ions is observed due to the presence of the MOT. Interestingly, though there exist several pathways to the formation of $^{40}$Ca$^{+}$ -- $\textit{e.g.}$ photo-ionization of an atom in the 5$^1$P$_1$ state, populated via the repumping channel, by a 423~nm photon -- an ion signal is still observed even when the LQT is operated such that the charge-to-mass ratio of $^{40}$Ca$^{+}$ is not trapped. The mass of this anomalous ion species is measured by the LQT stability parameters, a comparison of the relative CEM time-of-flight to that of ions with known mass ($\textit{e.g.}$ $^{40}$Ca$^{+}$) and, ultimately, by resonant excitation of the secular trap frequency as shown in Fig. \ref{secularscan}.  The excitation measurement is performed by applying a small (relative to the trapping potentials) oscillating voltage to one of the trap electrodes, inducing substantial heating that results in trap loss when the excitation is resonant with the mass-specific secular frequency of the ion motion.  The excitation spectrum is taken by allowing the LQT to load with ions for 1.9~s at each excitation frequency before detection with the CEM.  The ratio of detected ion number with excitation to detected ion number without excitation is plotted in Fig. \ref{secularscan}.   The trap loss resonance (and its second harmonic) is observed as the reduction in ion signal near $70 \pm 2$~kHz ($140 \pm 5$~kHz), in good agreement with the predicted radial secular frequency of 76~kHz for 80~amu, confirming the production of $^{40}$Ca$_2^+$ molecular ions.  The splitting of the fundamental peak is likely due to trap imperfections such as the asymmetric axial electrode placement, asymmetric voltages, and imperfect machining.

\begin{figure}
\resizebox{1\columnwidth}{!}{
    \includegraphics{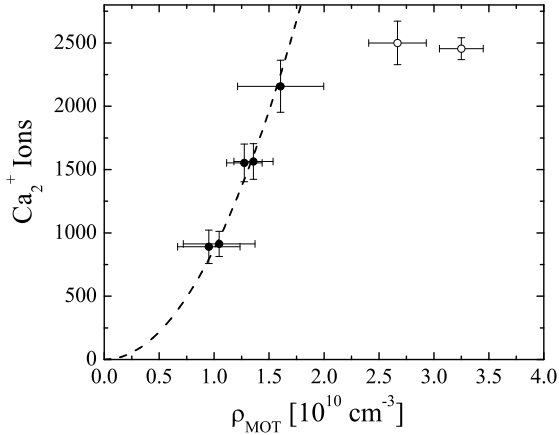}
}  \caption{Ion number measured as a function of the MOT density determined by absorption imaging.  A quadratic trend (with a best fit shown to guide the eye) is observed until trap saturation occurs at higher densities (shown in the points with open circles).  \label{SaturationDensityGraph}}
\end{figure}

 The LQT loading dynamics of the dimer ion are shown in Fig. \ref{Ca2+LoadingUnloading}.  For the data in Fig. \ref{Ca2+LoadingUnloading}(a), the LQT is first loaded with dimer ions from the MOT for a designated amount of time after which the ion number is measured by the CEM.  As seen in this panel, the characteristic loading time is found to be $\sim 1$~s and typically leads to an equilibrium ion number of $\sim 3000$ ions.  To measure the competing effects which lead to this equilibrium, we measure the trap lifetime as shown in Fig. \ref{Ca2+LoadingUnloading}(b). For this measurement, we allow the trap to load to saturation and then switch off the MOT at t = 0~s, stopping molecular ion production. After a designated time, we detect the ion number via the CEM.  In this way, we find the molecular ion lifetime in the LQT to be $\sim 90$~s. Since inelastic processes between the ions and neutral atoms~\cite{footnote1} do not change the number of dimer ions in the trap, the loading time is not affected by the presence of the MOT.  Therefore, the large disparity between the molecular ion lifetime and its characteristic loading time shows that the equilibrium condition is not set by competition between the natural trap loss mechanisms and the dimer ion production rate.  Instead, equilibrium is determined by a maximum ion density that sets a trap capacity at which the Coulomb repulsion between the ions already in the trap and the freshly formed ions is large enough to overwhelm the trapping potential \cite{Douglas2005}. Further evidence to support this conclusion comes from measurements at different MOT parameters with different characteristic loading times which reach the same approximate equilibrium ion number, as well as measurements counting $^{40}$Ca$^{+}$ and $^{174}$Yb$^{+}$ loaded by photo-ionization and laser ablation, respectively.

To confirm that the dimer ion is produced by collisions within the MOT and not by other means, such as collisions of MOT atoms with hot Ca from the getter, we measure the production of molecular ions as a function of MOT density, as shown in Fig. \ref{SaturationDensityGraph}.  For the data in this figure we detect the dimer ion number after 1.9~s of loading at various densities. We vary the density by controlling the intensity of the 672~nm repump laser; the density at each point is determined by absorption imaging. Quadratic dependence is found for sufficiently small densities, but trap saturation, due to the space charge effect, leads to constant ion number at higher densities. The dependence of the molecular ion production rate at smaller densities is indicative of a two-body process as would be expected for PAI between MOT atoms.

\begin{figure}
\resizebox{1\columnwidth}{!}{
    \includegraphics{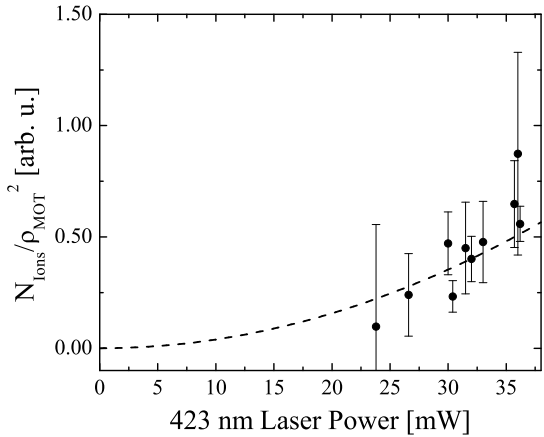}%
} \caption{Normalized production rate as a function of overall 423~nm laser power incident upon the MOT.  A quadratic fit is shown as the dashed line.\label{powerplot}}
\end{figure}

To determine the number of photons required for the PAI, we conduct a similar experiment as a function of overall 423~nm laser light power with results shown in Fig.~\ref{powerplot}.  Since the position and density of the MOT are sensitive to the exact power balance between the deceleration and trapping beams, there is higher systematic uncertainty in this measurement.   However, we observe a trend that is consistent with quadractic dependence, drawn as a dashed line to guide the eye in Fig.~\ref{powerplot}.  Thus, we conclude that the $^{40}$Ca$_2^+$ molecular ions are most likely produced by a two-body, two-photon production mechanism as detailed in what follows. 
\begin{figure}
\resizebox{1\columnwidth}{!}{ \includegraphics{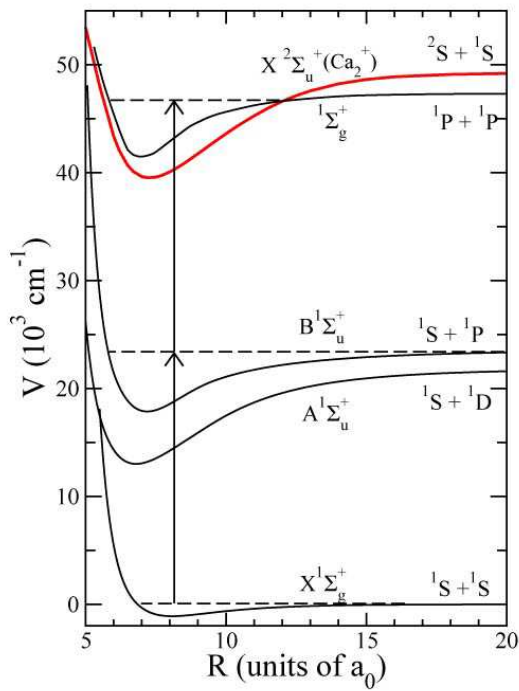}} 
\caption{Molecular potentials of $^{40}$Ca$_2$ (black curves) and $^{40}$Ca$_2^+$
(red curve).  Two vertical arrows indicate the energy carried in the
423~nm MOT photons.  We observe that the ground state potential of the molecular ion is in the same energy region as the potentials dissociating to two $^1$P $^{40}$Ca atoms.
\label{CaPotlSvetlana}}
\end{figure}
Because space charge effects dominate the LQT loading dynamics, the molecular ion production rate cannot simply be analyzed in terms of the competing rate equation model typically employed in trap loading dynamics. Therefore, in order to estimate this rate,  we postulate that the number of Ca$_2^+$ molecular ions, $N$, in the ion trap as a function of time is given as:
\begin{eqnarray}
\frac{dN}{dt} &=& \int\eta(\vec{r}) \beta(\vec{r}) \rho_{\rm{Ca}}(\vec{r})^2 \left( 1-\frac{N}{N_o}  \right) d^{3}\vec{r}\nonumber \\
 &=&  \bar{\eta}  \bar{\beta} V \bar{\rho}_{\rm{Ca}}^2 \left( 1-\frac{N}{N_o} \right), \nonumber \label{eqn1}
\end{eqnarray}
where the average production rate is characterized by $\bar{\beta}$, the two-body intensity dependent rate coefficient; $\bar{\eta}$ is the average of a spatially varying capture efficiency which quantifies the probability of successfully trapping an ion formed at a particular location; $\bar{\rho}_{\rm{Ca}}^2$ and V are the average squared MOT density and volume, respectively. The last factor is an $\textit{ad hoc}$ model to take into account the trap saturation at the observed maximal ion number $N_o \sim 3000 \pm 300$.  This model allows us to use full loading curve measurements to set a lower bound on the production rate in a way that is equivalent to performing a model independent linear fit to only the very early ($N \sim 0, t\sim 0$) portion of the loading curve, where space charge effects should be minimal. The rate coefficient from the fit, assuming perfect trap capture efficiency ($\bar{\eta} \rightarrow 1$), to the loading curve shown in Fig. \ref{Ca2+LoadingUnloading} is found to be $\bar{\beta} = 2 \pm 1 \times 10^{-15}$~cm$^{3}$~Hz.  This rate coefficient should only be interpreted as a lower bound to the true rate constant since the capture efficiency is taken to be its maximal value.  Interestingly, this rate coefficient is comparable to observations in alkali systems \cite{Band-1992, Bagnato1993} when scaled to the relatively low intensities found in this experiment.   This bound is independent of any assumptions about the specific mechanism leading to the PAI event.


In order to determine the specific PAI pathway for
molecular ion production,  molecular potential curves for the neutral $^{40}$Ca$_2 $ and ionic $^{40}$Ca$_2^+$ molecules are presented in Fig.~\ref{CaPotlSvetlana}. The potentials of the X$^1\Sigma_g^+$ and B$^1\Sigma_u^+$ states of $^{40}$Ca$_2$ are constructed
from spectroscopic data in Refs.~\cite{Allard2003,Allard2004}.
Other potentials are determined using non-relativistic configuration-interaction molecular-orbital restricted-active-space (MOL-RAS-CI) calculations.  From these molecular potentials it is clear that there are several energetically allowed pathways for associative ionization that must be considered:
\begin{eqnarray}
{\rm Ca} + {\rm Ca}^{+} &\rightarrow& {\rm Ca}_{2}^{+} + \gamma \label{ion collision} \\
{\rm Ca(P)} + {\rm Ca(S)} +\gamma &\rightarrow& {\rm Ca}_{2}^{+} + e^{-} \label{s-p} \\
{\rm Ca(P)} + {\rm Ca(P)} &\rightarrow& {\rm Ca}_{2}^{+} + e^{-} \label{p-p}  \\
{\rm Ca(D)} + {\rm Ca(S)} + \gamma &\rightarrow&  {\rm Ca}_{2}^{+} + e^{-}  \label{d-s} \\
{\rm Ca(S)} + {\rm Ca(S)} + 2\gamma &\rightarrow&   {\rm Ca}_{2}^{+} + e^{-} \label{s-s-multi}
\end{eqnarray}

The first pathway (\ref{ion collision}) is the radiative association of the atomic ion with its neutral partner.  To investigate this reaction, we trap large amounts of $^{40}$Ca$^+$ and monitor the production of $^{40}$Ca$_2^+$ in the presence of the MOT. We see no evidence of this production mechanism and conclude that radiative association cannot be the dominant pathway.

The  next two pathways are $\textit{single step}$ processes in which one of the collision partners resides in an excited state before the collision reaches the Condon point, where a free-to-bound transition is possible. The excited collision pathways (\ref{s-p},\ref{p-p}) are ruled out because the duration of the cold collision is longer than the excited 4$^1$P$_1$ state lifetime.  As discussed in \cite{Band-1992}, excited atoms are unlikely to be in close enough proximity to undergo association since the 423~nm laser driving the transition is far detuned due to the inter-atomic interaction even at distances large compared to the Condon length.  

\begin{figure}
\resizebox{1\columnwidth}{!}{
    \includegraphics{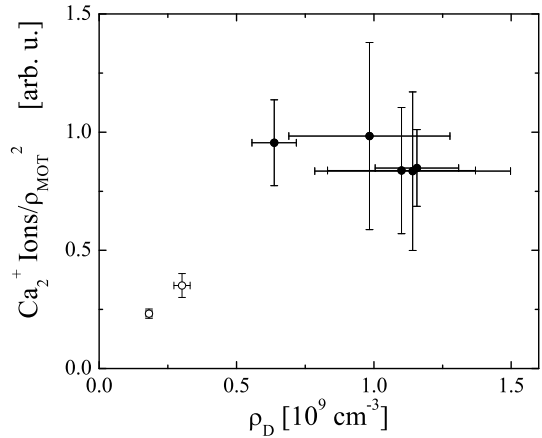}
}  \caption{Ion number normalized by the square of the MOT density as a function of 3$^{1}$D$_{2}$ state density.  Note the open circle points correspond to those in Fig. \ref{SaturationDensityGraph} at which trap capacity is reached. \label{rhodplot}}
\end{figure}

The other excited collision pathway (\ref{d-s}) cannot be ruled out due to collision duration argument \cite{Band-1992} because the $3^1D_2$~Ca state is relatively long lived, with a radiative lifetime $\sim 1$~ms, and could survive for the duration of a collision event.  To quantify this production mechanism, we must consider the dependence of the dimer ion production on the partial density of  $3^1$D$_2$~Ca in the MOT.  In steady state, the density of 3$^{1}$D$_{2}$  state Ca can be determined as $\rho_{D} = \left(\Gamma_{\rm{PD}} \rho_{\rm{Ca}} \rho_{ee}\right)/\left(\Gamma_{\rm{rad}} + \Gamma_{\rm{KE}} \right)$ where $\rho_{\rm{Ca}} \rho_{ee}$ is the density of atoms in the 4$^{1}$P$_1$ state, $\Gamma_{\rm{PD}}$ is the radiative decay rate for the 4$^{1}$P$_1$~$\rightarrow$ 3$^1$D$_2$ transition, $\Gamma_{\rm{rad}}$ is the total radiative loss rate from the 3$^1$D$_2$ state, and $\Gamma_{\rm{KE}}$ is the ballistic escape rate from the MOT volume.  The ballistic escape rate is determined to be $1.4\pm0.2$~kHz by fitting to the cloud volume evolution measured during free expansion of the MOT atoms.  The total radiative loss rate is determined by solving a rate equation model for the effect of the repump laser on the 3$^{1}$D$_{2}$ state population that includes the 5$^{1}$P$_1$ and 4$^{1}$S$_0$ states as well as the losses into the 3$^{3}$P$_{2,1}$ states.  Using this system of rate equations, the partial density of 3$^{1}$D$_{2}$ $^{40}$Ca is determined for each data point in Fig. \ref{SaturationDensityGraph}.  The result, after dividing out the overall quadratic density dependence, is plotted in Fig. \ref{rhodplot}.  Despite the fact the density of 3$^{1}$D$_{2}$ Ca varies by over factor of two, the rate constant is unchanged.  We note the two points in open circles correspond to the points in Fig. \ref{SaturationDensityGraph} where the ion number is space charge limited, and should be ignored.  Therefore it appears that the 3$^{1}$D$_{2}$ is unimportant for the observed PAI and thus pathway (\ref{s-p}) is ruled out.  

Therefore, we conclude that the two-photon PAI process (\ref{s-s-multi}) is responsible for the observed production of $^{40}$Ca$^{+}_{2}$ molecular ions. This formation proceeds by the excitation scheme sketched in Fig.~\ref{CaPotlSvetlana}, where colliding neutral atoms are photo-associated into an excited neutral molecule state, which is subsequently excited to the ground molecular ion state or a further excited, auto-ionizing neutral molecule state. We note that this pathway agrees well with the conclusions of Ref. \cite{Weiner-1999}.

In summary, we have observed the formation of $^{40}$Ca$^{+}_{2}$ molecular ions and found a lower limit of its production rate constant at $\bar{\beta} \geq 2 \pm 1 \times 10^{-15}$~cm$^{3}$~Hz. From the dependences of the production rate on MOT parameters and $\textit{ab initio}$ calculations, we have determined the production mechanism is most likely two-step, two-photon photo-associative ionization out of the ground state molecular potential.  We produce approximately 3000 molecular ions, limited only by the space charge capacity of the ion trap. As this technique does not require a separate photo-association laser, it could find use as a simple, robust method for producing ultracold, state-selected molecular ions.

This work was supported by NSF grant No. PHY-1005453, ARO grant No. W911NF-10-1-0505 and AFOSR grant.



\bibliography{ca2testbib,Svetlana}

\begin{thebibliography}{23}
\bibitem[{foo()}]{footnote1}
\bibinfo{note}{An inelastic process leading to the production of heavier
  molecular ions, such as $^{40}$Ca$_{3}^{+}$, could lead to dimer ion loss.
  However, a direct search for the appearance of heavier ions found no evidence
  for their production.}
\end{thebibliography}
\end{document}